\documentclass{article}
\usepackage{arxiv}
\usepackage[T1]{fontenc}
\usepackage{graphicx}
\usepackage{enumitem}
\usepackage{amsfonts}
\usepackage{subcaption}
\usepackage{comment}
\usepackage{booktabs}
\usepackage{pdfpages}
\usepackage{tabularx, multirow}
\usepackage{xcolor}
\usepackage{orcidlink} 
\usepackage{wrapfig}
\usepackage{algorithm}
\usepackage{algpseudocode}
\usepackage{todonotes}
\usepackage[many]{tcolorbox}
\usepackage{tablefootnote}

\usepackage{amsmath}

\usepackage{pifont} % cmark, xmark
\newcommand{\cmark}{\ding{51}}%
\newcommand{\xmark}{\ding{55}}%
\newcommand{\code}[1]{\texttt{#1}}

\usepackage[belowskip=2pt,aboveskip=2pt]{caption}
\makeatletter
\newcommand\footnoteref[1]{\protected@xdef\@thefnmark{\ref{#1}}\@footnotemark}
\makeatother

\usepackage{mwe}
\usepackage{wrapfig}

\begin{document}
\sloppy

\renewcommand{\headeright}{}
\renewcommand{\undertitle}{(Accepted at IEEE PRIME 2026)}
\renewcommand{\shorttitle}{HW-SW Co-Design for F16 On-Device Training on RISC-V Single-Core}

\title{Hardware-Software Co-Design for Float16 On-Device Training on RISC-V Single-Core}

\date{}

\author{ {Benjamin Hubinet$^{1,2}$, Pierre-Alain Moëllic$^{1,2}$, Olivier Savry$^{2}$, Olivier Potin$^{3}$, Jean-Baptiste Rigaud$^{3}$} \\
	$^{1}$ CEA-Leti, Mines Saint-Etienne, Equipe Commune SAS, F-13541 Gardanne, France \\
	$^{2}$ Univ. Grenoble Alpes, CEA-Leti, F-38000 Grenoble, France\\
	\texttt{\{name\}.\{surname\}@cea.fr} \\
	$^{3}$ Mines Saint-Etienne, CEA-Leti, Centre CMP, Equipe commune SAS, F-13541 Gardanne, France\\
	\texttt{olivier.potin@emse.fr, rigaud@emse.fr} \\
}

\maketitle

\begin{abstract}
By leveraging standard RISC-V extensions, namely Zfh (scalar float16) and Zvfh (vector float16), this work proposes an open-source framework to enable complete on-device training on resource-constrained RISC-V single-core. Our approach allows memory footprint reduction by about 50\% as compared to using float32 and with minimal model performance degradation. We also facilitate transfer learning and fine-tuning scenarios by incorporating layer-freezing capabilities. Our work builds onto AIfES, an open-source, modular and generic DNN training and inference framework for embedded systems that can be extended with custom hardware-specific functions. The benefits of float16 is further emphasized by outlining the low area overhead of Zfh on a RV64GC super-scalar out-of-order FPGA softcore (+1.15\% LUT6 and +0.05\% FF at 175MHz). Finally, we discuss the architecture of a Zvfh implementation within the same RISC-V core.

\keywords{On-Device Training \and Half-Precision Floating-Point\and RISC-V\and Deep Neural Networks\and Resource-Constrained Devices\and FPGA.}
\end{abstract}

\setcounter{footnote}{0} 

\section{Introduction}
Given the compute and memory extensiveness of On-Device Training (ODT) of Deep Neural Network (DNN) models, various algorithmic strategies have been explored, % in recent years, 
notably partial model training \cite{cai2020tinytl}, quantized/hybrid training \cite{ribeiro2025decentor} or both \cite{deutel2024device}. Besides restricted fine-tuning scenarios, these hybrid training approaches struggle to support full-fledged training or the use of a batch size greater than 1. These limitations primarily stem from using float32 only for the computationally expensive backward pass and relying on quantized operations for the forward pass, which ultimately leads to significant learning instability. In addition, while memory constraints may prevent on-device training of a float32 DNN, smaller floating-point formats like float16 may be better suited. In that sense, the RISC-V scalar float16 extension Zfh and its vector counterpart Zvfh can be leveraged to reduce the memory footprint over float32 and yield SIMD execution from Zvfh. Consequently, to address the lack of open-source ODT frameworks optimized for RISC-V single-core supporting Zfh and/or Zvfh, \textbf{we propose an easy-to-use open-source library\footnote{\url{https://gitlab.emse.fr/securityml/frameworks/aifes4riscv}} which allows PyTorch/Tensorflow models to be deployed and fully trained on RISC-V single-core featuring Zfh/Zvfh support.}

In Table \ref{tab:comparaison_sota}, we compare different available ODT frameworks. % with our contribution. 
Although being a RISC-V-targeted open-source framework that offers full scalar float16 ODT capabilities, PULP-Trainlib \cite{nadalini2022pulp} is specific to RISC-V multi-core platforms and limited to a batch size of 1. In contrast, AIfES \cite{wulfert2024aifes} is a full ODT capable framework supporting batch sizes greater than 1, but lacks both float16 and RISC-V-specific support.

\begin{table}[h!]
       \caption{Comparison of our framework with others}
       \centering
       \begin{tabular}{c c c c c c}
               \toprule
                                          & RISC-V  & Full ODT   & Float16    & Batch       & Open \\
               Work                       & support & capability & support    & size        & source \\
               \toprule
               \cite{ribeiro2025decentor} & \cmark & \xmark & \xmark & 1                 & \xmark \\
               %% MCUNetv3: \cite{lin2022device}       & \xmark & \xmark & \xmark & 1                 & \cmark \\
               \cite{deutel2024device}    & \xmark & \cmark & \xmark & $\geq 1$          & \xmark \\
               \cite{nadalini2022pulp}    & \cmark & \cmark & \cmark & 1                 & \cmark \\
               \cite{wulfert2024aifes}    & \xmark & \cmark & \xmark & $\geq 1$          & \cmark \\
               \textbf{Ours}              & \cmark & \cmark & \cmark & \textbf{$\geq$ 1} & \cmark \\
               \bottomrule
               \vspace{-.8cm}
       \end{tabular}
       \label{tab:comparaison_sota}
\end{table}

To yield the best performance, optimized DNN software libraries are often combined to specialized hardware units, such as systolic-array-based matrix multiplication (GeMM) accelerators, consisting of Multipliy-and-Accumulate (MAC) processing elements arranged in an array to maximize spatial data reuse. Albeit speeding-up the execution over a purely sequential software GeMM implementation, current GeMM accelerators lack standardized RISC-V ISA support. Thus, their integration into system-on-chip either translates as memory-mapped peripherals \cite{tortorella2022redmule} or using custom ISA extensions \cite{genc2019gemmini,kamaleldin2026procon}, which regardless requires a dedicated and non-standard software stack. More versatile and RISC-V-backed accelerators also exist, namely Vector Processing Units (VPU) for which the RISC-V Vector extension (RVV) guarantees code portability across different implementations. As RVV can be divided into subsets, various degrees of open-source implementations have been proposed. For instance, RISC-V² \cite{patsidis2020risc} which focuses on vector integer arithmetic acceleration, Vicuna2 \cite{jones2025vicuna2} which brings full Zvfh support to a scalar in-order core in a minimalist way RVV-wise, or Ara2 \cite{perotti2024ara2} which completely implements RVV 1.0 and can scale up to multi-core vector processors. Amongst the existing open-source implementations, none propose strict Zvfh support to a super-scalar out-of-order core. 

To this end, we introduce the architecture of our Zvfh implementation within the NaxRiscv: an FPGA-optimized configurable super-scalar out-of-order RISC-V core. As a first step towards achieving a Zvfh-compliant VPU, we also share the implementation details and cost of its scalar counterpart Zfh.%}

\section{Extending AIfES for float16 ODT on RISC-V}

\subsection{Framework architecture}
We base our work on AIfES \cite{wulfert2024aifes}, % as a seed for our work, 
as its modular nature allows it to be extended with hardware-specialized implementations, including Zfh/Zvfh hand-optimized math primitives. Although auto-vectorization is supported by modern compilers like GCC and LLVM, recent works \cite{van2024muriscv,mahmoudi2025tinyml} have shown that manual vectorization still outperforms auto-vectorized scalar kernels. Hence, manually vectorized kernels were used, namely the NMSIS-DSP\footnote{\url{https://doc.nucleisys.com/nmsis/dsp/index.html}} open-source library from Nuclei System Technology, inspired from ARM CMSIS-DSP. Despite NMSIS-DSP was originally developed for Nuclei-based processors, it implements maths primitives optimized for standard extensions, i.e. RISC-V Zfh and Zvfh. Precisely, Zvfh ensures code portability by its vector-length agnostic nature.
\begin{figure}[h!]
    \centering
    \includegraphics[width=.7\columnwidth]{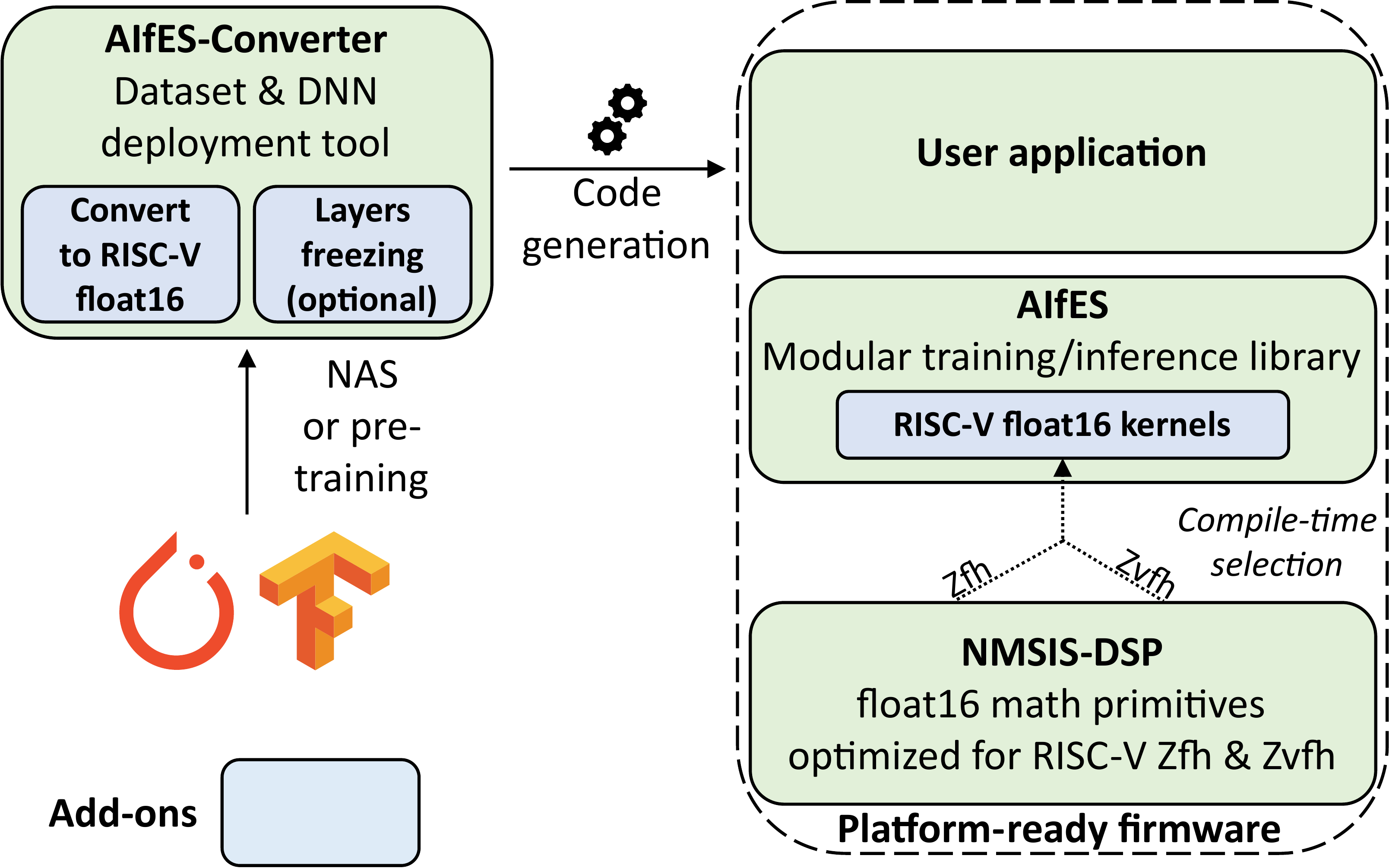}
    \caption{Overall pipeline based on AIfES with our add-ons supporting Zfh and Zvfh optimized DNN kernels.}
    \vspace{-.4cm}
    \label{fig:aifes_modification}
\end{figure}
Thus, NMSIS-DSP primitives are suitable for any RISC-V single-core implementing Zvfh. All the changes that we have brought to AIfES are outlined Fig. \ref{fig:aifes_modification}. To ease prototyping, AIfES-Converter, a tool capable of converting PyTorch or TensorFlow models into AIfES models in C was also enhanced. The latter improvements allow the deployment of models incorporating the added RISC-V kernels. Furthermore, we provide users with the option to freeze specific layers to support transfer learning or fine-tuning scenarios.

\subsection{Framework evaluation}
\begin{figure*}[t!]
    \centering
    \includegraphics[width=.9\textwidth]{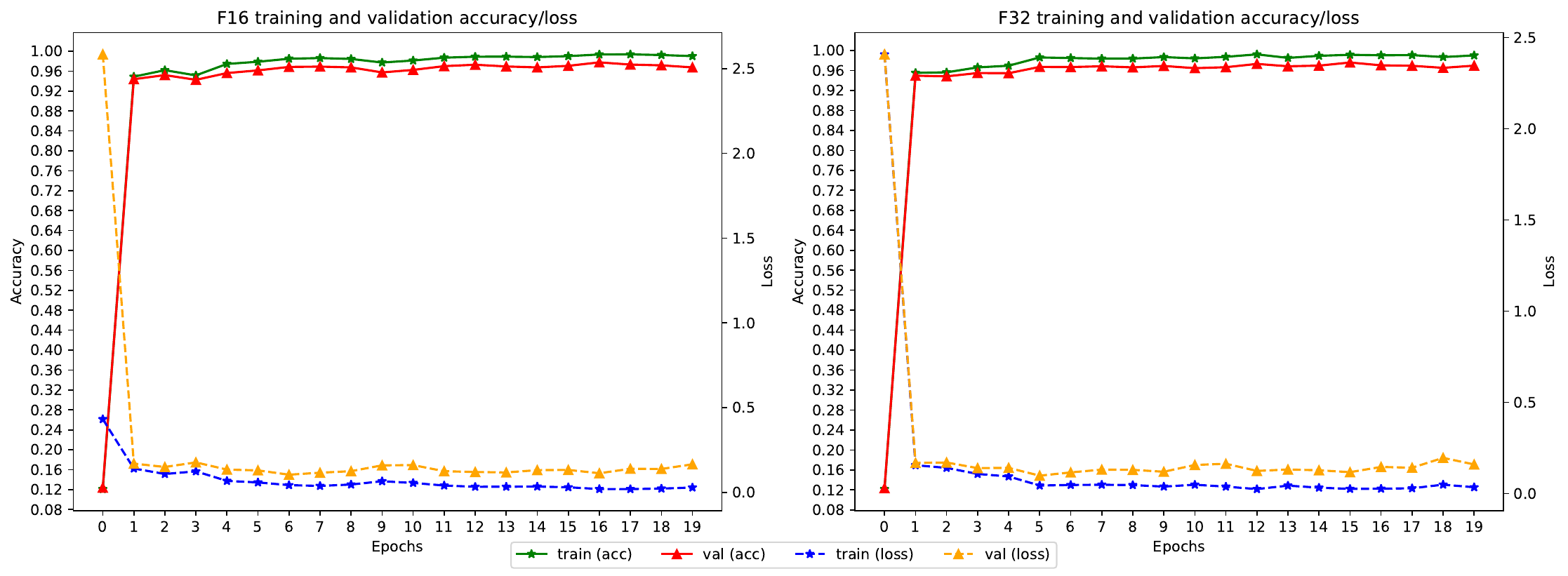}
    \caption{Training a MLP from scratch on MNIST, on 20 epochs, using our RISC-V Zvfh optimized kernels (left) versus the default float32 kernels (right)}
    \label{fig:mnist_aifes_training_f16_vs_f32}
\end{figure*}% Pour changer du RISC-V SUMMIT, au besoin, j'ai aussi une version FashionMNIST

We trained\footnote{batch size: 128, optimizer: adam, loss: cross-entropy, lr: 1e-3} a 5-layer MLP (512-256-128-64-10 units with ReLU as activation) on MNIST, using the RISC-V ISA Simulator. As shown in Fig.~\ref{fig:mnist_aifes_training_f16_vs_f32}, we reach very similar performance to float32 training, according to the loss and accuracy evolution. In addition, further training experiments were led using the same architecture but with varying datasets, learning rates and optimizers. Table \ref{tab:training_results} confirms that using float16 only results in slight accuracy difference as compared to float32.

In a transfer learning scenario, with a 3-layer MLP (32-16-10) pretrained on MNIST and fine-tuned on EMNIST, we measure the memory benefits of float16 training in Table \ref{tab:implementation_results}. "Model" refers to the model parameters memory (storage memory), while "Gradients" refers to the gradients, optimizer and activation memory (work memory). We reach a flat 50\% reduction for the parameters. Nonetheless, memory usage can be further brought down when using the SGD optimizer instead of Adam, and by only training the last layer.
\begin{table}[h!]
       \caption{Performance comparison of float16 versus float32 training}
       \centering
       \begin{tabular}{c c c c c}
               \toprule
               Dataset & Learning & Optimizer & Float16        & Float32 \\
                       & rate     &           & val. acc. (\%) & val. acc. (\%) \\
               \toprule
               MNIST        & 1e-3  & SGD   & 97.29  & 96.88 \\
               %\midrule
               MNIST        & 1e-3  & Adam  & 96.78  & 97.32 \\
               %\midrule
               FashionMNIST & 1e-3  & SGD   & 86.85  & 87.53 \\
               %\midrule
               FashionMNIST & 1e-4  & Adam  & 88.25  & 88.54  \\
               \bottomrule
               \vspace{-.7cm}
       \end{tabular}
       \label{tab:training_results}
\end{table}
\begin{table}[h!]
       \caption{Global memory usage for different configurations of a MLP}
       \centering
       \begin{tabular}{c c cc}
               & & \multicolumn{2}{c}{Gradients (kB)} \\
               \cmidrule(r){3-4}
               Frozen layer id & Model (kB) & SGD & Adam \\
               \toprule
               (\textbf{f16}) $\emptyset$ & 50.43  & 101.43 & 202.75 \\
               (\textbf{f32}) $\emptyset$ & 100.85 & 202.60 & 404.77 \\
               \midrule
               (\textbf{f16}) 1-2 & 50.43  & 52.74  & 154.06 \\
               (\textbf{f32}) 1-2 & 100.85 & 105.23 & 307.40 \\
               \bottomrule
       \end{tabular}
       \label{tab:implementation_results}
\end{table}

\section{Bringing float16 support to the NaxRiscv}
\subsection{NaxRiscv overview\label{subsection:naxriscv_intro}}
\begin{figure}[h!]%t
    \centering
    %\hspace*{-5mm}
    \includegraphics[width=.7\columnwidth]{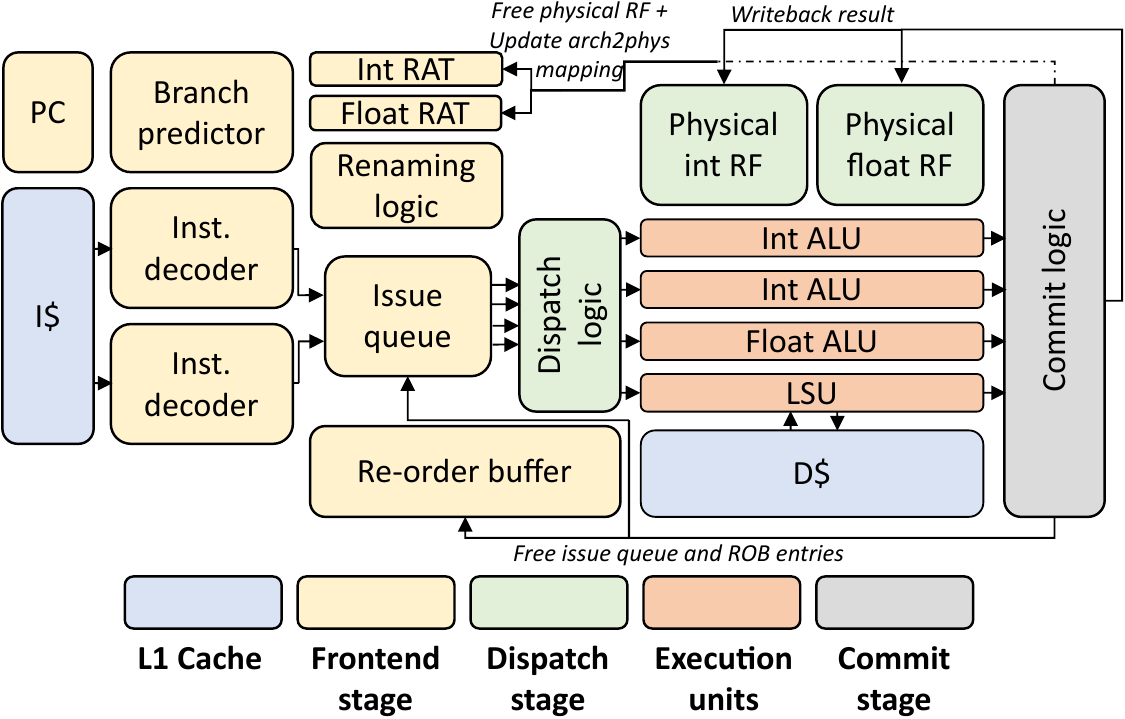}
    \caption{High-level overview of the NaxRiscv}
    \vspace{-.4cm}
    \label{fig:naxriscv_top}
\end{figure}

For this work, we considered the NaxRiscv, an open-source configurable RV64GC super-scalar out-of-order core targeting FPGA. The NaxRiscv is written in SpinalHDL, a meta hardware description language which provides zero-cost hardware abstractions to easily write generic hardware descriptions. Besides providing abstraction to digital designers, SpinalHDL seemlessly integrates with existing EDA tools as it is in effect a Verilog and VHDL generator.
As outlined in the high-level block diagram Fig. \ref{fig:naxriscv_top}, we chose a dual-issue NaxRiscv configuration comprising a 32 entries issue queue, two integer Arithmetic Logic Unit (ALU), one Floating-Point Unit (FPU), one Load-Store Unit (LSU), a 64 entries re-order buffer, a Gshare branch predictor and two L1 16kB 4 ways caches for respectively instructions and data.

\subsection{Scalar float16 (Zfh) support}
In order to achieve full Zvfh compliance, a float16 capable FPU is needed. First, the FPU is an actual building block of a Zvfh-compatible VPU. Moreover, the Zvfh specification mandates full scalar float16 compatiblity, that is Zfh support. Therefore, we extended the NaxRiscv instructions decoder, LSU and FPU to accomodate the 36 Zfh instructions. Apart from decoding these new instructions, little change is required on the LSU and FPU side as the latter already support both scalar float32 and float64 operations. Precisely, the narrow float32 are loaded and stored as 64-bits float values through Nan-Boxing, a process of binary extension by left-padding with ones. At compute time, as the FPU comprises functional units with a 64-bits wide data-path, float32 operands are unpacked to their IEEE standard format, and are then extended to float64 by means of exponent biasing and logical left-shifting. As a result, we applied the same policy to obtain float16 support.

The FPGA resources overhead to a standard RV64GC NaxRiscv is considerably low, namely +1.15\% LUT6 and +0.05\% FF at 175MHz on a Xilinx XC7K325T-2FFG900C target\footnote{We used Vivado 2021.1 for this implementation}. Additionally, no maximum operating frequency degradation (175MHz) is observed when enabling the Zfh extension.

\subsection{Vector float16 (Zvfh) support}
\begin{figure}[t!]%t
    \centering
    %\hspace*{-5mm}
    \includegraphics[width=.7\columnwidth]{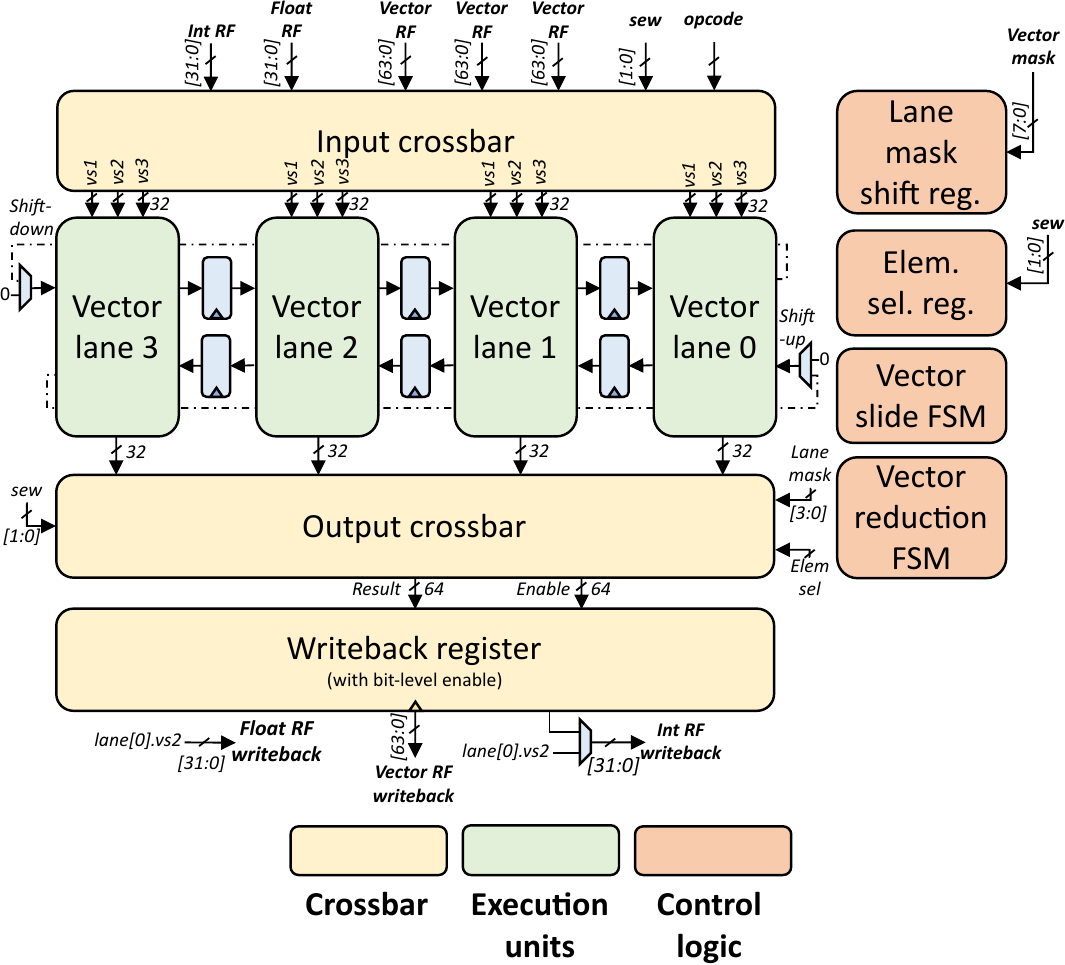}
    \caption{High-level overview of the  
    VPU}
    \vspace{-.0175cm}
    \label{fig:vpu_top}
\end{figure}

Aside from Zfh support, no other side extensions are needed for a full-fledged Zvfh implementation. As a consequence, we discuss the architectural choices for our VPU, depicted Fig. \ref{fig:vpu_top}. For the VPU itself, we adopted a lane architecture, that is a set of FPUs and integer ALUs processing vector elements in parallel. The latter are fed by a crossbar that routes source operands elements based on their format, the vector mask and the instruction type. Once done, lane results are packed together using a second crossbar. Depending on the operation and whether there are more elements to be processed than the number of lanes, processed elements are held in a writeback register until the instruction fully completes. Finally, the full result is written back to the correct register file.

As we mostly emphasize on vector float16 acceleration and as XLEN=64-bits in the case of a RV64GC NaxRiscv, we chose a vector length of VLEN=64-bits, such as to reach a balance between parallelism and hardware reuse. Specifically, with four lanes, 64-bits wide vectors provide a raw 4$\times$ speed-up over scalar execution for float16 operands. Moreover, a VLEN of 64-bits allows merging both the vector and the float physical register file and part of their respective register allocation table, thus limiting the hardware overhead. Unlike other VPUs, we chose to tightly integrate our into the NaxRiscv as the SpinalHDL's plugins system provides design flexibility in a ad-hoc fashion. Thus the existing instruction decoder is extended to support Zvfh instructions.

\begin{figure}[h!]%t
    \centering
    \includegraphics[width=0.5\columnwidth]{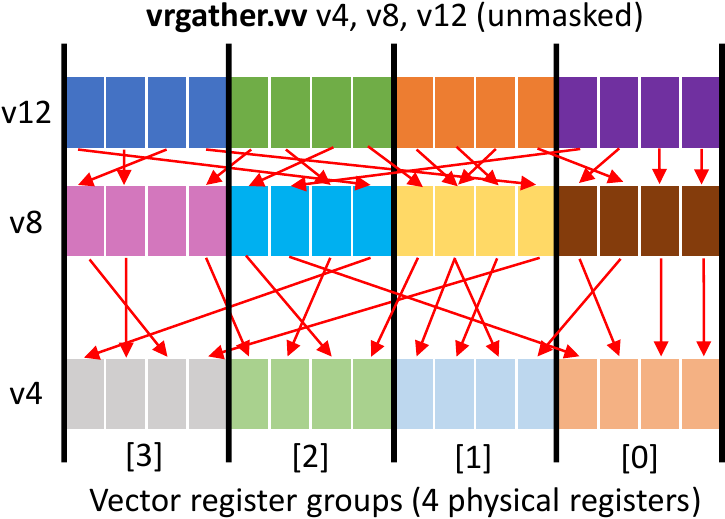}
    \caption{\code{vrgather} example when LMUL=4}
    \label{fig:vrgather_lmul_4_example}
\end{figure}

To allow for longer vectors, the RISC-V Zvfh specification introduced register grouping, namely regrouping several physical vector registers (2, 4 or 8) under the same architectural vector\footnote{Regrouped vector registers are also named \textit{vector register groups}}. We chose to handle this Zvfh property by breaking vector instructions into micro-operations during instruction issue/renaming, as to produce instructions with single physical register operands and foster their out-of-order execution. However, we do not generate micro-ops for all vector instructions. For instance, \code{vrgather.vv vd, vs2, vs1} performs an indirection on vector register groups, namely it copies into \code{vd} the elements within \code{vs2} at indices given by \code{vs1}. Consequently, as shown Fig. \ref{fig:vrgather_lmul_4_example}, each physical destination register (\code{v4}) may receive elements from multiple physical source registers (\code{v8}) depending on indices (\code{v12}). Hence, slicing such an instruction would result in $(LMUL)^2$ micro-operations as all the possible destination-source registers relations must be accounted for.

\section{Conclusion}
This work demonstrates that full float16 ODT is possible on resource-constrained RISC-V single-core supporting standard extensions (i.e. Zfh and/or Zvfh), while requiring about 50\% less memory than float32. Furthermore, as a result of enhancement to AIfES-Converter, our contribution makes fine-tuning and transfer learning a possibility. We also show that by its FPGA resources friendliness on a RV64GC NaxRiscv, Zfh alone can be a good candidate to enable float16 ODT at the hardware level. Future work will focus on evaluating the performance gains brought by vector kernels on our VPU-enhanced NaxRiscv FPGA target, to complement with current results focused on validating the ODT pipeline through the RISC-V ISA Simulator and evaluating the implementation cost of RISC-V Zfh into our target.

\section*{Acknowledgements}
This work is supported by the French ANR in the Investissements d’avenir (ANR-10-AIRT-05, irtnanoelec), AI.MMUNITY program. Works were provided with computing and storage resources by GENCI (grant AD011011932) on the Jean Zay supercomputer.

\bibliographystyle{splncs04}
\bibliography{bibliography}

\end{document}